\documentstyle[prl,aps,epsf,multicol]{revtex}
\begin{document}
\draft
\title{Plasticity and memory effects in the vortex solid phase of twinned YBa$_2$Cu$_3$O$_7$
single crystals}
\author{S. O. Valenzuela and V. Bekeris}
\address{Laboratorio de Bajas Temperaturas, Departamento de
F\'{\i}sica, Universidad Nacional de Buenos Aires, Pabell\'on I, Ciudad
Universitaria, 1428 Buenos Aires, Argentina}
\date{Submitted 9 December 1999}
\maketitle
\begin{abstract}
We report on marked memory effects in the vortex system of twinned
YBa$_2$Cu$_3$O$_7$ single crystals observed in ac susceptibility
measurements. We show that the vortex system can be trapped in
different metastable states with variable degree of order arising
in response to different system histories. The pressure exerted by
the oscillating ac field assists the vortex system in ordering,
locally reducing the critical current density in the penetrated
outer zone of the sample. The robustness of the ordered and
disordered states together with the spatial profile of the
critical current density lead to the observed memory effects.
\end{abstract}
\begin{multicols}{2}
\narrowtext
Continuous efforts have been made to understand the remarkably
rich variety of liquid and solid phases in high temperature
superconductors \cite{blat}. A subject that has recently attracted
much interest is the connection between these thermodynamic phases
and the driven motion of vortices, in particular concerning the
presence of topological defects (like dislocations) and the
evolution of the spatial order of the vortex structure (VS) at
different driving forces
\cite{Kosh,yaron,marl,Matsu,hend1,flavio,abu,hend2,rav}. In
systems containing random pinning, theoretical \cite{Kosh} and
experimental \cite{yaron,Matsu,hend1,flavio} results have shown
that, at the depinning transition, the VS undergoes plastic flow
in which neighboring parts of the flux lattice move at different
velocities thereby disordering the VS.

Changes in the volume in which vortices remain correlated may
modify the critical current density $J_{\mathrm{c}}$ and lead to a
thermomagnetic history dependence in the transport and magnetic
properties of the superconductor in a way reminiscent of other
disordered systems such as spin glasses \cite{let}. History
effects recently observed in conventional low-$T_{\mathrm{c}}$
superconductors \cite{hend1,hend2,rav,kupf} were attributed to
plastic deformations of the VS, however, little is known about the
exact mechanism involved in these phenomena. In YBa$_2$Cu$_3$O$_7$
(YBCO), the importance of plasticity has been revealed through
magnetic \cite{abu,zies,kup,kokk} and transport \cite{fend}
measurements below the melting transition, though detailed history
effects studies have not been performed up to now.

    In this work we report on thermomagnetic history effects in the solid
vortex phase of pure twinned YBCO single crystals by measuring the
ac susceptibility with the ac field parallel to the {\it c} axis
of the sample. ac susceptibility is a sensitive tool to detect
changes in $J_{\mathrm{c}}$ and therefore, in the translational
correlation length of the VS. Our results show that the VS may be
trapped in different metastable states depending on its
thermomagnetic history. For example, if the sample is cooled from
above $T_{\mathrm{c}}$ with no applied ac field, the VS is trapped
in a more disordered state than when the ac field is turned on
during the cooling process. In addition, we find evidence that the
cyclic pressure exerted by the ac field induces a dynamical
reordering of the VS in the penetrated outer zone of the sample
that persists when the ac field is turned off and, as a result,
different parts of the sample may be in different pinning regimes.
The resulting spatial variation of $J_{\mathrm{c}}$ leads to a
strong history dependence of the ac response and accounts for the
observed memory effects. The character of the reordering induced
by the ac field seems to be related to the flow of dislocations in
a similar way as in ordinary solids under cyclic stress
\cite{hertz}.

Magnetic field orientation relative to the twin boundaries (TB) is
another ingredient that determines the overall ac response. The
effective strength of pinning at the TB's can be tuned by rotating
the applied dc field out of the twin planes. At small angles,
where pinning by TB's is expected to be more effective, history
effects are weak. At larger angles, however, the influence of TB's
diminishes and then pronounced history effects are observed.

\begin{figure}
\epsfclipon
\narrowtext

\vskip 0mm
\epsfxsize=.9
\hsize
\centerline{\vbox{
\epsffile{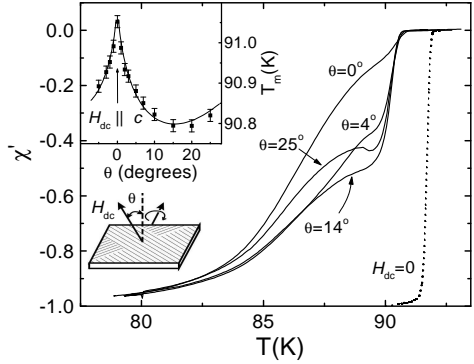} }}
\vskip -4mm

\caption{Temperature dependence of the screening $\chi'$ for the
twinned YBCO single crystal for different orientations of the dc
field relative to the {\it c} axis of the sample.
$H_{\mathrm{dc}}$=3KOe, $h_{\mathrm{ac}}$=2Oe and $f$=10.22kHz.
The upper inset shows the angular dependence of $T_{\mathrm{m}}$.
The line is a guide to the eye. The lower inset shows the
orientation of the magnetic field and twin boundaries.}
\label{fig1}
\end{figure}

    Global ac susceptibility measurements with the mutual
inductance technique were carried out in two twinned single
crystals of YBCO. We present the data obtained with one of them
(dimensions $0.56 \times 0.6 \times 0.02 ~ \mathrm{mm}^{3}$). The
crystal has a $T_{\mathrm{c}}$ of 92 K at zero dc field
($h_{\mathrm{ac}}$=1 Oe) and a transition width of 0.3 K (10-90\%
criterion). Polarized light microscopy revealed that the crystal
has three definite groups of twins oriented $45^{\circ}$ from the
crystal edge as shown in the inset of Fig. \ref{fig1}.
Susceptibility data were recorded under different angles $\theta$
between the applied field and the {\it c} axis. The axis of
rotation is shown in the inset of Fig. \ref{fig1} and it was
chosen so that the field can be rotated out of all twin boundary
planes simultaneously.

We begin by describing briefly the angular dependence of the ac
susceptibility and then we will focus our discussion on the memory
effects. Fig. \ref{fig1} shows the real component of the ac
susceptibility, $\chi'$, for four values of $\theta$. The data
were obtained while decreasing the temperature in the usual field
cooled procedure at a rate of 0.2 K/min, with $H_{\mathrm{dc}}$ =
3 kOe, and a superimposed ac field of amplitude $h_{\mathrm{ac}}$
= 2 Oe and frequency $f$ = 10.22 kHz.

The diamagnetic screening, $\chi'$, presents a dramatic evolution
as $\theta$ is increased from $\theta$ = 0$^\circ$
($H_{\mathrm{dc}}\parallel c$). The pronounced changes are a
consequence of the effective pinning strength of the TB's. Fig.
\ref{fig1} shows that, when the dc field is tilted from the TB's
direction, a sharp onset in the susceptibility develops. A sharp
onset in the susceptibility was observed in untwinned YBCO single
crystals \cite{giapin,brac} and was demonstrated to coincide
closely \cite{brac} with a sharp resistivity drop that is
generally accepted as a fingerprint of a melting transition
between a vortex liquid phase and a vortex solid phase with
long-range order \cite{hugo,kw}. We identify the step-like onset
of $\chi$' as the melting temperature, $T_{\mathrm{m}}$. The
absence at small angles of the sharp onset in $\chi$' (Fig.
\ref{fig1}) suggests that the first order melting transition is
suppressed by the TB's. The nature of the transition for
$H_{\mathrm{dc}}\parallel c$ is generally accepted to be a second
order phase transition to a Bose-Glass state \cite{nel}. A peak at
$\theta = 0^{\circ}$ in the angular dependence of
$T_{\mathrm{m}}$, as seen in the inset of Fig. \ref{fig1}, has
been related to this transition \cite{kw,nel}.

The angular dependence of the ac susceptibility in twinned samples
is still a matter of debate. The observed minimum in the shielding
at low temperatures and $\theta = 0^{\circ}$ has been recently
explained by vortex channeling along TB's \cite{ed}. According to
Ref. \cite{ed,ous}, as the field is rotated out of the TB's, the
channeling is partially suppressed. This would explain the initial
increase in $\chi$' at small angles (see Fig. \ref{fig1}).
However, above a threshold angle, $\theta_{\mathrm{k}} \sim
14^{\circ}$, a new reduction in $\chi$' is observed (Fig.
\ref{fig1}, $\theta = 25^{\circ}$). One reason may be that, for
these angles, the influence of TB's vanishes and a more ordered VS
can form \cite{zies} leading to a reduction in $J_{\mathrm{c}}$.

\begin{figure}
\narrowtext
\vskip 0mm
\epsfxsize=.8
\hsize
\centerline{\vbox{
\epsffile{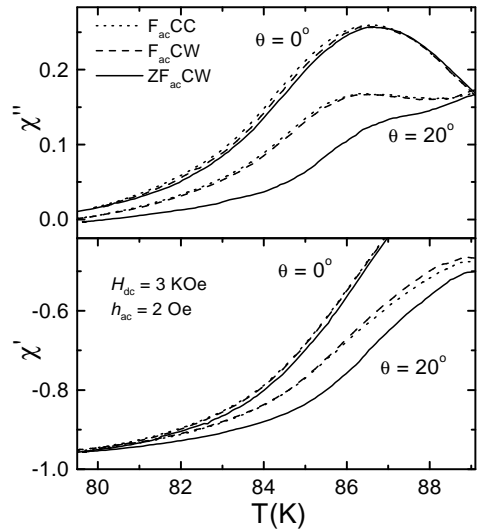} }}
\vskip -12mm
\caption{$\chi'(T)$ and $\chi''(T)$ for different magnetic
histories. Dotted curves where obtained on cooling,
F$_{\mathrm{ac}}$CC. Dashed curves on warming after measuring the
dotted curves, F$_{\mathrm{ac}}$CW. Solid curves on warming after
cooling with no ac field applied, ZF$_{\mathrm{ac}}$CW.
$H_{\mathrm{dc}}$=3kOe, $h_{\mathrm{ac}}$=2Oe and $f$=10.22kHz.}
\label{fig2}
\end{figure}

\begin{figure}
\narrowtext \vskip 0mm
\epsfxsize=.8 \hsize \centerline{\vbox{ \epsffile{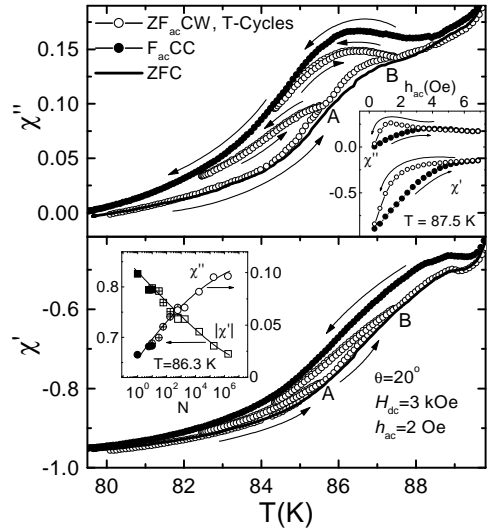} }}
\vskip -12mm
\caption{Memory effects. Main panel: $\chi'(T)$ and $\chi''(T)$
measured during temperature cycles. Measurements start at $T \sim
80$K after ZF$_{\mathrm{ac}}$C (open circles). Also shown are the
F$_{\mathrm{ac}}$CC (solid circles) and ZFC (solid line) curves.
Arrows indicate the direction of the temperature sweep.
$\theta$=$20^{\circ}$, $h_{\mathrm{ac}}$=2Oe. Top inset: Cycles in
ac field amplitude. $T$=87.5K. Arrows indicate the direction of
the ac field sweep. Bottom inset: $|\chi'|$ (squares) and $\chi''$
(circles) vs. $N$. $h_{\mathrm{ac}}$=1Oe, $f$=10.22kHz. Black
symbols where measured after ordering with $h_{\mathrm{ac}}$=2Oe
and $f$=0.1Hz, + in center symbols with $f$=1Hz and open symbols
(from left to right) with 10Hz, 30Hz, 300Hz, 3kHz and 30kHz. Lines
are guides to the eye.} \label{fig3}
\end{figure}

We turn now to the memory effects. The main results of our
investigation are summarized in Fig. \ref{fig2} and \ref{fig3}.
The measurements in Fig. \ref{fig2} were performed by varying $T$
at fixed $H_{\mathrm{dc}}$, $h_{\mathrm{ac}}$ and $\theta$. Dotted
curves were obtained on cooling (C) as the ones in Fig.
\ref{fig1}, while solid and dashed curves were obtained on warming
(W). The difference between solid and dashed curves is the way the
sample was cooled prior to the measurements. Dashed curves were
performed after cooling from $T > T_{\mathrm{c}}$ with applied ac
field (F$_{\mathrm{ac}}$CW), {\it e.g.} after measuring the dotted
curves. Solid curves were also obtained after cooling from $T >
T_{\mathrm{c}}$ but with $h_{\mathrm{ac}}$ = 0
(ZF$_{\mathrm{ac}}$CW). It is apparent that when cooling the
sample with no applied ac field the vortex system solidifies in a
more strongly disordered and pinned state (with a higher effective
critical current density $J_{\mathrm{c}}^{\mathrm{dis}}$), as can
be inferred from the enhanced shielding and the reduced
dissipation.

The angular dependence of the thermomagnetic history exhibits
interesting features. At small $\theta$, the three curves are very
similar, though a closer look shows that the maximum shielding
(and minimum dissipation) corresponds to the ZF$_{\mathrm{ac}}$CW
case. As $\theta$ is increased, but kept below
$\theta_{\mathrm{k}}$, the thermomagnetic history becomes more and
more relevant. For angles near $\theta_{\mathrm{k}}$, the
importance of the history rapidly increases and the
ZF$_{\mathrm{ac}}$CW case strongly separates from the other two
curves (see Fig. \ref{fig2}). The relative variation between the
measured $\chi$' for different sample histories is seen to be as
high as 20\%. The magnitude of the history dependence at angles
beyond $\theta_{\mathrm{k}}$ suggests that the first order melting
transition and the translational order of the VS are key factors
to explain this behavior.

The results presented above can be understood in terms of a
dynamical reordering of the VS caused by the shaking movement
induced by the applied ac field during the cooling process in the
F$_{\mathrm{ac}}$CC and the F$_{\mathrm{ac}}$CW cases. Due to this
dynamical ordering the correlation length of the VS grows and
$J_{\mathrm{c}}$ diminishes ($J_{\mathrm{c}} =
J_{\mathrm{c}}^{\mathrm{ord}} < J_{\mathrm{c}}^{\mathrm{dis}}$).
Note that when $H_{\mathrm{dc}} \parallel c$ a long range ordered
structure is unlikely to form as TB's prevent vortices from
occupying positions favored by their mutual interaction. This
seems to be the case even when the ac field is applied manifesting
in the faint history effects shown in Fig. \ref{fig2}. However,
when the field is tilted from the twins their influence weakens
and history effects become more evident suggesting the dynamical
ordering of the VS by the ac field.

This reordering or {\it annealing} of the VS occurs in the
penetrated outer zone of the sample which depends on the ac field
amplitude and the temperature (as $J_{\mathrm{c}}$ is temperature
dependent). If the VS is initially disordered ({\it e.g.}
ZF$_{\mathrm{ac}}$C), an increase in $T$ or $h_{\mathrm{ac}}$ will
order the VS as the ac field front progresses towards the center
of the sample. The VS at the inner region will remain disordered
if the condition $J < J_{\mathrm{c}}^{\mathrm{dis}}$ is satisfied
at {\it all} times and an elastic Campbell-like regime applies
with most of the vortices pinned \cite{bra}. From this follows
that the {\it spatial profile} of $J_{\mathrm{c}}$ is determined
by the {\it history} of the sample. If the sample is cooled from
$T > T_{\mathrm{c}}$, it will be measured the smallest shielding
for the applied ac field (F$_{\mathrm{ac}}$CC and
F$_{\mathrm{ac}}$CW cases), as observed in Fig. \ref{fig2}.

If the above reasoning is correct, memory effects both in $T$ and
$h_{\mathrm{ac}}$ should be observed because the inner disordered
state is able to sustain a higher current without vortex movement,
thereby enhancing the shielding $|\chi'|$ and reducing the
dissipation $\chi''$ in the sample. Moreover, the existence of the
disordered region should become more evident in $\chi$ when the ac
flux front is near its boundary. These memory effects are clearly
depicted in Fig. \ref{fig3}. Starting at $T \sim 80$ K with a
disordered state (ZF$_{\mathrm{ac}}$C) we measured the
susceptibility while increasing temperature
(ZF$_{\mathrm{ac}}$CW). As $T$ increases, $J_{\mathrm{c}}$
decreases and the ac field front penetrates further into the
sample ordering the VS. If warming is stopped and the temperature
is lowered (point $A$) the measured susceptibility shows a
hysteretic behavior as the outer part of the sample is now
ordered. Furthermore, when the sample is driven to low enough
temperatures the measured susceptibility tends to that obtained in
the F$_{\mathrm{ac}}$CC procedure because the ac field is unable
to sense the inner disordered region that was never reached by the
ac field front. If now the temperature is increased once again,
the susceptibility closely follows the last cooling curve because
the order-disorder profile was not changed during the cooling
process. As expected, beyond point $A$ the ac response matches the
ZF$_{\mathrm{ac}}$CW case. The procedure was repeated at point $B$
where an equivalent description can be made. It is worth noting
that the long term memory (our experiments take more than 1 h)
indicates that both the disordered and the ordered metastable
states are very robust.

Analogous cycles in $h_{\mathrm{ac}}$ that corroborate the above
explanation can be performed starting with $h_{\mathrm{ac}}$ = 0
after ZF$_{\mathrm{ac}}$C (top inset of Fig. \ref{fig3}). Note
that for ac field amplitudes higher than 5.5 Oe no hysteretic
behavior is observed in neither $\chi'$ nor $\chi''$. In this
case, the ac field has penetrated in the whole sample suppressing
the disordered region. As the VS has been fully annealed, the
system loses memory of the highest ac field that was applied.

We also studied what happens when the sample is zero field cooled
(both ac and dc)(ZFC). The dc field rate was 50 Oe/s. In these
measurements, the ac field is turned on after the dc field has
reached its final value. The warm up curves obtained after
preparing the system at $T \sim 80$ K are also contained in Fig.
\ref{fig3} (solid line). The results are close to the disordered
ZF$_{\mathrm{ac}}$CW ones. We interpret this in terms of disorder
yielded by plastic motion of vortices. It is well established from
Bitter decoration \cite{flavio}, SANS \cite{yaron}, noise
\cite{marl} and transport experiments \cite{hend1,hend2} that when
a current near $J_{\mathrm{c}}$ drives the flux lattice a
disordered plastic motion occurs. On the other hand, there is
experimental \cite{abu} and theoretical \cite{feig} evidence
suggesting the existence of a dislocation mediated plastic creep
of the vortex structure that would be analogous to the diffusive
motion of dislocations in solids \cite{hertz}. In our experiment,
when the magnetic field is applied, vortices can start penetrating
only when the induced current $J$ is of the same order of
$J_{\mathrm{c}}$. In this situation, a small dispersion in the
pinning strength will destroy the long range order generating a
high density of defects. As this plastic motion proceeds, the
screening current decreases and the lattice will be unable to
reorder as detected when we apply the ac field \cite{sov}.

While from the above discussion it is clear that the ZFC case will
correspond to a disordered state, one can then ask on the
character of the VS annealing when an ac field is applied. The
mechanism involved in this phenomenon may be analogous to the flow
and rearrangement of dislocations that leads to the softening of
hard atomic solids under cyclic stress \cite{hend2,hertz}. The
bottom inset of Fig. \ref{fig3} shows evidence of this
cycle-dependent softening in the VS. Starting from a
ZF$_{\mathrm{ac}}$C disordered state, the sample is cycled with a
large ac field to anneal the VS. After $N$ cycles the ac field is
turned off and the state of the VS is sensed by measuring the ac
susceptibility with a smaller probe ac field at 10 kHz. This
procedure is repeated for each point in the figure. As discussed
above, the degree of exclusion of the probe and therefore the
susceptibility are a function of the critical current of the
sample at the moment the probe is applied ($J_{\mathrm{c}}(N)$).
The cycle-dependent behavior of the susceptibility nicely
demonstrates that $J_{\mathrm{c}}$ is also cycle-dependent.

In conclusion, we have presented susceptibility measurements on
twinned YBCO single crystals. We find that the vortex system can
be trapped in different metastable states as a consequence of
different thermomagnetic histories. When measuring susceptibility,
the oscillating applied field assists the vortex structure in
ordering, locally reducing the critical current density. As a
result, different parts of the sample can be in different pinning
regimes. The robustness of these states and the associated spatial
variation of the critical current density manifest in strong
memory effects both in temperature and ac field. The angular
dependence of these effects is consistent with an increase of the
correlation length of the VS when the dc field is rotated out of
the twin planes.

We expect that similar effects will be present in transport
measurements because, in most cases, the applied (ac) current will
not flow homogeneously inside the sample and will force a field
redistribution in a similar manner as when applying an external ac
field.

We acknowledge E. Rodr\'{\i}guez and H. Safar for a critical reading of
the manuscript. This research was supported by UBACyT TX-90,
CONICET PID N$^{\circ}$ 4634 and Fundaci\'on Sauber\'an.

\end{multicols}
\end{document}